\documentclass[12pt]{iopart}
\usepackage{iopams}
\usepackage{epsfig}   
\usepackage{graphics}
\usepackage{lscape}
\usepackage{rotating}

\begin{document}

\title[Quantifying the impact of systematic uncertainties in supernova
  cosmology]{Quantifying systematic uncertainties in supernova cosmology}

\author{Jakob Nordin$^1$, Ariel Goobar$^1$ and Jakob J\"onsson$^{1,2}$}

\address{$^1$ Department of Physics, Stockholm University, Albanova
         University Center, \\
         S--106 91 Stockholm, Sweden}
\address{$^2$ University of Oxford Astrophysics, Denys Wilkinson
         Building, Keble Road, \\ Oxford OX1 3RH, UK } 

\begin{abstract}
Observations of Type Ia supernovae used to map the expansion history of the 
Universe suffer from systematic uncertainties that need to be propagated 
into the estimates of cosmological parameters. We propose an iterative
Monte-Carlo simulation and cosmology fitting technique (SMOCK) 
to investigate the impact of sources of error upon fits of the dark energy equation of state. This approach is especially useful to track the impact 
of non-Gaussian, correlated effects, e.g.~reddening correction errors, 
brightness evolution of the supernovae, K-corrections, gravitational
lensing, etc.
While the tool is primarily aimed for studies and optimization of future
instruments, we use the ``Gold'' data-set in  Riess et al.~(2007) to
show examples of potential systematic uncertainties that could exceed
the quoted statistical uncertainties. 
\end{abstract}

\eads{\mailto{nordin@physto.se}, \mailto{ariel@physto.se}, \mailto{jacke@astro.ox.ac.uk}}




\section{Introduction}
The direct evidence for dark energy, which were obtained through 
observations of Type Ia
supernovae (SN~Ia) in the 1990s, marked the beginning of a new era in
observational
cosmology~\cite{goo95,garnavich98,riess98,schmidt98,perl99}. 
Significantly
improved SN~Ia data sets have since then been reported in
e.g.~\cite{knop03,tonry03,barris04,riess04}. These data sets have been
followed by the impressive results from the ongoing dedicated SN~Ia
surveys: SNLS~\cite{astier06} and 
ESSENCE~\cite{wood}.
An important
recent compilation of published SNe~Ia, including e.g.~the supernovae from 
the first year of SNLS, but also adding several high redshift 
($z \gtrsim 1$) SNe~Ia 
from the GOODS survey, is reported in \cite{riess06}.

As the SN~Ia sample sizes keep growing at an increased rate, the
relative importance of statistical uncertainties is steadily
decreasing, which brings systematic uncertainties to focus.  It is
thus of key importance to estimate the impact of systematic
uncertainties as we enter the era of ``precision cosmology''.

In this work we propose a Monte Carlo simulation approach to quantify the
propagated effect of known (or suspected) systematic effects in supernova
cosmology. The emphasis here is to quantify the degradation of 
the precision to estimate cosmological parameters due to
systematic errors, possibly correlated in arbitrary groupings of supernovae: in redshift
bins, by locations in the sky, observational instruments, time 
of observations, etc.

The current study includes investigations of sources of
error from: uncertainties in the extinction by dust, K-corrections,
redshift uncertainties, gravitational lensing, Malmquist bias, 
possible brightness
evolution of the ``calibrated'' standard candle (e.g. metallicity 
effects), spectral template differences,
and miss-calibrations between the low and the high redshift 
supernova sample.
Systematic effects caused by bulk motion have been specifically
targeted in a number of recent studies, e.g.~\cite{neill07}, and were
not specifically included in this study. However, some general
systematic effects, e.g. drift of estimated peak brightness
at low-z, would mimic the main effect expected 
from peculiar velocities.

The proposed method, implemented in the SMOCK package and presented
in~\S\ref{sec:method}, allows us to quantify the sensitivity
of cosmological results, obtained from real or simulated data,  
to specific (multiple) systematic effects. For example, we can
quantify at which level a systematic effect gives rise to a bias
exceeding the statistical uncertainty.
The tool could hence be used to define requirements
of any future mission aimed at improving our knowledge of the dark
sector using SNe~Ia.
As an example of how this technique can be used we have performed a dedicated
study of the ``Gold'' sample presented in Reference~\cite{riess06}. 
This study
demonstrates how cosmological parameter fits are affected by 
a number of different systematic errors.

In~\S\ref{sec:snmag} the model of the Universe which we have used 
and the cosmological parameters are
introduced. Our methodology for studying the effects of systematic
uncertainties is described in~\S\ref{sec:method}. 
In~\S\ref{sec:sysquant} we show how
the changes in the cosmological results due to systematic
uncertainties can be quantified and the trends parametrized.
A number of
different systematic effects, their implementation, and their effects 
on the Gold set are the topics of~\S\ref{sec:syseff}. 
Some observations regarding systematic effects and the
Gold set are presented in~\S\ref{sec:disc}. The paper is concluded by
some implications for current and future SN~Ia surveys in~\S\ref{sec:concl}. 

Throughout the paper units are used where the speed of light is unity,
i.e.~$c=1$.

\section{SN~Ia magnitudes and cosmological parameters} \label{sec:snmag}
In this paper we investigate the accuracy to which the 
cosmological
parameters $\vec{\theta}=(\Omega_M,\Omega_X,w_0,w_a)$ can be
determined.
These parameters describe the matter density, $\Omega_M$, the
dark energy density, $\Omega_X$, and the dark energy equation
of state parameter, which is parametrized by $w(z)=w_0+w_a z/(1+z)$.

The apparent magnitude in an arbitray observer filter $Y$, $m_Y$, of a SN~Ia at 
redshift $z$ is given by:
\begin{equation}
    m_Y(\vec{\theta},{\cal M}_X,z)=
    {\cal M}_X+5\log_{10}\left[d'_L(\vec{\theta},z)\right] + K_{XY}(z) +
         A_{XY}(z),
\end{equation}
where
\begin{equation}
{\cal M}_X=25+M_X-5\log_{10}H_0,
\label{eq:Mscript}
\end{equation}
is a ``nuisance'' parameter depending on the Hubble parameter, $H_0$,
and the brightness-shape corrected absolute magnitude, $M_X$, 
of the supernova in restframe
broadband filter $X$. The $H_0$-\emph{independent} luminosity distance
is denoted by
$d'_L\equiv H_0\,d_L$. 
$K_{XY}$ corresponds to the 
K-corrections involved in the transformation 
between magnitudes in the restframe $X$-band and the observed
magnitudes in the $Y$-band, as 
defined in Reference~\cite{kcorr}.
$A_{XY}$ is a general color correction which could include, for
example, corrections
for host galaxy extinction in the restframe $X$-band ($Y$-band 
in the observer frame) and 
intrinsic color dispersion.

The $H_0$-independent luminosity distance $d'_L$ is given by
\begin{eqnarray}
    d'_L&=&(1+z)\left\{
    \begin{array}{ll}
      \frac{1}{\sqrt{-\Omega_K}}\sin(\sqrt{-\Omega_K}\,I)  &
      \Omega_k<0\\
      I  & \Omega_K=0\\
      \frac{1}{\sqrt{\Omega_K}}\sinh(\sqrt{\Omega_K}\,I)  &
      \Omega_k>0\\
    \end{array}
    \right., \\
    \Omega_K&=&1-\Omega_M-\Omega_X,\\
    I&=&\int_0^z\,\frac{dz'}{H'(z')} ,\\
    H'(z)&=&H(z)/H_0=\nonumber\\
    &&\sqrt{(1+z)^3\,\Omega_{M}+
    f(z)\,\Omega_{X}+(1+z)^2\,\Omega_{K}}, \\
    f(z)&=&\exp\left[3\int_0^z\,dz'\,\frac{1+w(z')}{1+z'}\right] ,
\end{eqnarray}
where we consider, as already noted, an equation of state 
parametrized by
\begin{equation}
  w(z)=w_0+w_a z/(1 + z).
\end{equation}
We will also refer to simply '$w$' when a \emph{constant} 
($w_a=0$) equation
of state parameter for dark energy is considered, e.g.~in the figures
showing confidence level contours in the $(\Omega_M,w)$-plane.

In this article we focus on the implications of systematic
uncertainties on the estimation of the dark energy equation of state
parameter.  To obtain the results
presented here, a flat universe ($\Omega_K=0$) was assumed together
with constraints from the BAO peak in the SDSS sample of
luminous red galaxies~\cite{eis05}. There the parameter
\begin{equation}
A=\sqrt{\Omega_{m}}H'(z_1)^{-1/3}[\frac{1}{z_1}\int_0^{z_1}\frac{dz'}{H'(z')}]^{2/3}
\end{equation}
was measured to great precision at $z_1=0.35$: $A=0.469\pm0.017$. This was included as a prior.

Moreover, when showing results for
the $(w_0,w_a)$-plane we marginalize over the $\Omega_M$
parameter.  The method, presented in the next section, can of course
be applied to any other cosmological parameters using any combination
of priors.

\section{Methodology}
\label{sec:method}
The aim of this study is to investigate how different systematic
effects influence the cosmological parameters obtained
from fits on a set of SNe~Ia. In order to take both statistic
errors and systematic effects into account, we use Monte Carlo
simulations.
The method, in short, consists of the following steps:
\begin{enumerate}
\item The original data-set, consisting (minimally) of supernova
  redshifts and peak magnitudes, is used as a starting
  point\footnote{Instead of these derived properties, the supernovae
  could be described by their light curves observed in different
  wavelength bands. So far, our code does not calculate errors from
  modifications of individual measurement points.}.  Other properties,
  like excess color, can be included as well. A perturbed 
  synthetic data-set is
  generated by randomizing these quantities for each supernova
  according to either measurement uncertainties and/or assumed
  dispersion in the original data-set and the systematic
  effect(s). This is done taking into account possible correlations
  between supernovae.
\item Cosmological parameters, $\vec{\theta}$, are fitted to the
perturbed data-set.
\item The first two steps are repeated many times and the density of best fit
  values of cosmological parameters are used to trace out confidence 
level contours. From the relative location and shape of these
  contours it is possible to understand and to quantify the influence of
  systematic effects, both in terms of potential biases and enlarged 
  confidence level contours.
\end{enumerate}
Let us now elaborate on and motivate these steps.

The original data-set in step (i) could either be a simulated
data-set, corresponding to a known cosmology, or a real data-set for
which we would like to study the expected effects of systematic
errors.

When the perturbed synthetic data-sets are generated, all applicable
uncertainties and biases are used to compute the supernova redshifts
and peak magnitudes.  Since many sources of error may be included in a
single run, each contribution is co-added in the final perturbed
synthetic supernova data.  For example, the distance modulus to a
simulated supernova can be computed in several steps, each adding a
modification to the apparent magnitude. First, statistical
uncertainties are taken into account: a value of the peak magnitude
(e.g.~in rest-frame $B$-band) is drawn from a Gaussian distribution centered
around the original supernova brightness and with standard deviation
corresponding to the intrinsic brightness scatter together with (for
real data) the measurement error added in quadrature. Thereafter,
other perturbations are added. For example, a constant bias, mimicking
a calibration error, could be
added to all SNe~Ia in a particular redshift bin. 

Color errors or spectral template changes, and the
related reddening corrections, also result in offsets from the
originally derived distance modulus.  Redshift uncertainties are
generated from the probability distribution, if given by photometric
redshift error studies, or by a smaller Gaussian error, if
spectroscopic.  All of the above modifications can be applied to the
complete data set or just a subset (e.g.~only high redshift objects).
If only the intrinsic scatter, the measurement error or the sum added
in quadrature is used (and assumed to be Gaussian in magnitudes), the
standard analysis of the data-set is retrieved.

In this paper, such a
data-set will be referred to as a ``clean'' data-set.  Two advantages
with this method are that uncertainties and biases can be
\emph{non-Gaussian} and 
\emph{correlated}. 

Non-Gaussian error distributions can be studied since the measurements can be randomized according to any arbitrary distribution. Examples of such distributions are the dimming by dust and gravitational lensing magnification. Arbitrary correlations can be created through modifications of subsets of the data.

Any uncertainties in priors used should also be
incorporated according to this method. If the prior is Gaussian, 
a value of the prior is drawn from a Gaussian
distribution for each iteration.

In step (ii) the best fit cosmological parameters $\vec{\theta}$ for
 the 
synthetic data-set are
found using a maximum likelihood technique.

In this paper, a likelihood analysis taking
only Gaussian errors into account is performed. To find the best fit 
cosmological parameters to the synthetic data-set, the 
negative likelihood is minimized using the Powell
algorithm~\cite{powell} implemented in the SNALYS fitting routine within the
SNOC package~\cite{snoc}.  

\begin{figure}[hbtp]
\begin{center}
 \epsfig{figure=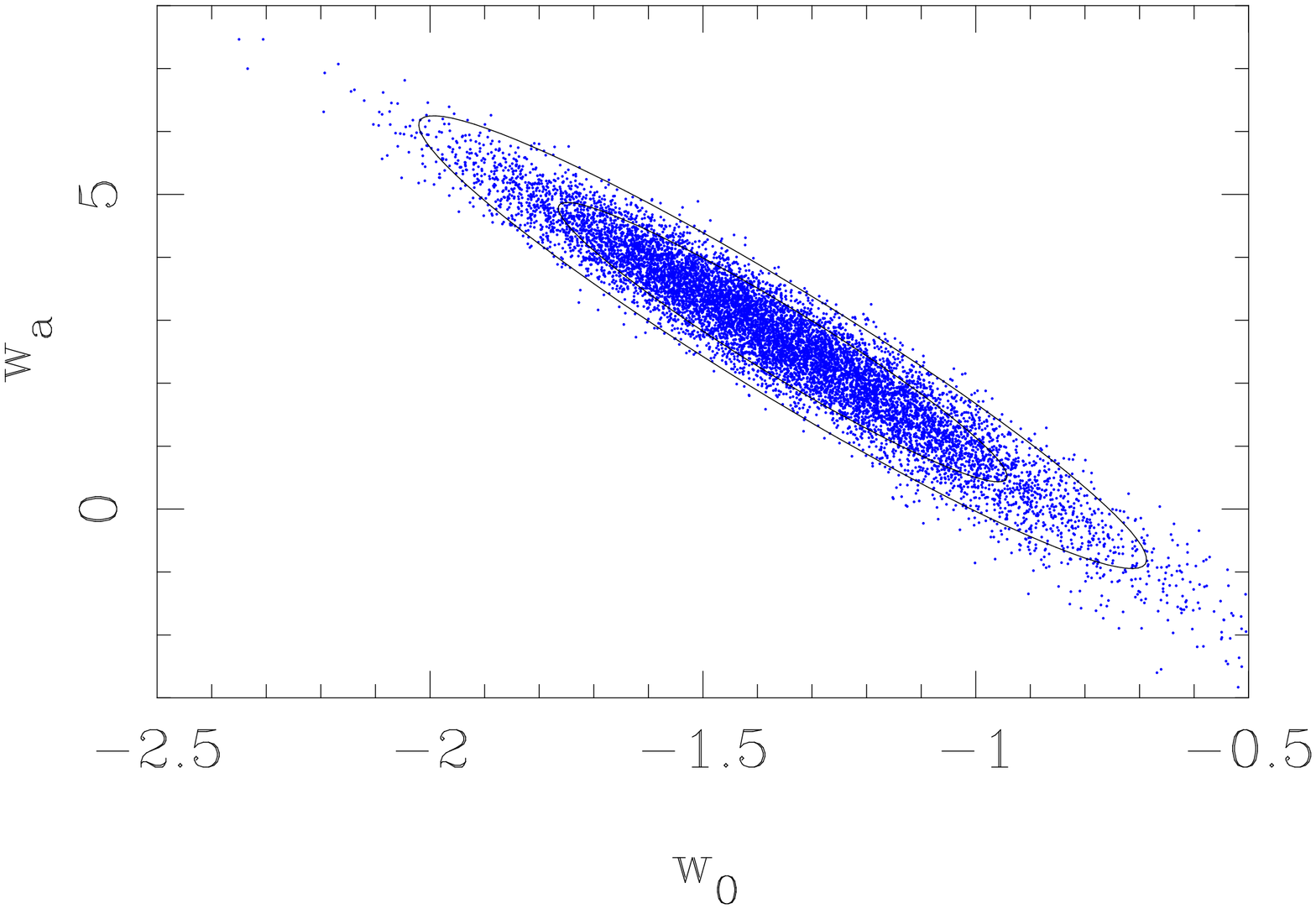, angle=-90,width=0.45\textwidth,bb=130 40 560 640}
 \epsfig{figure=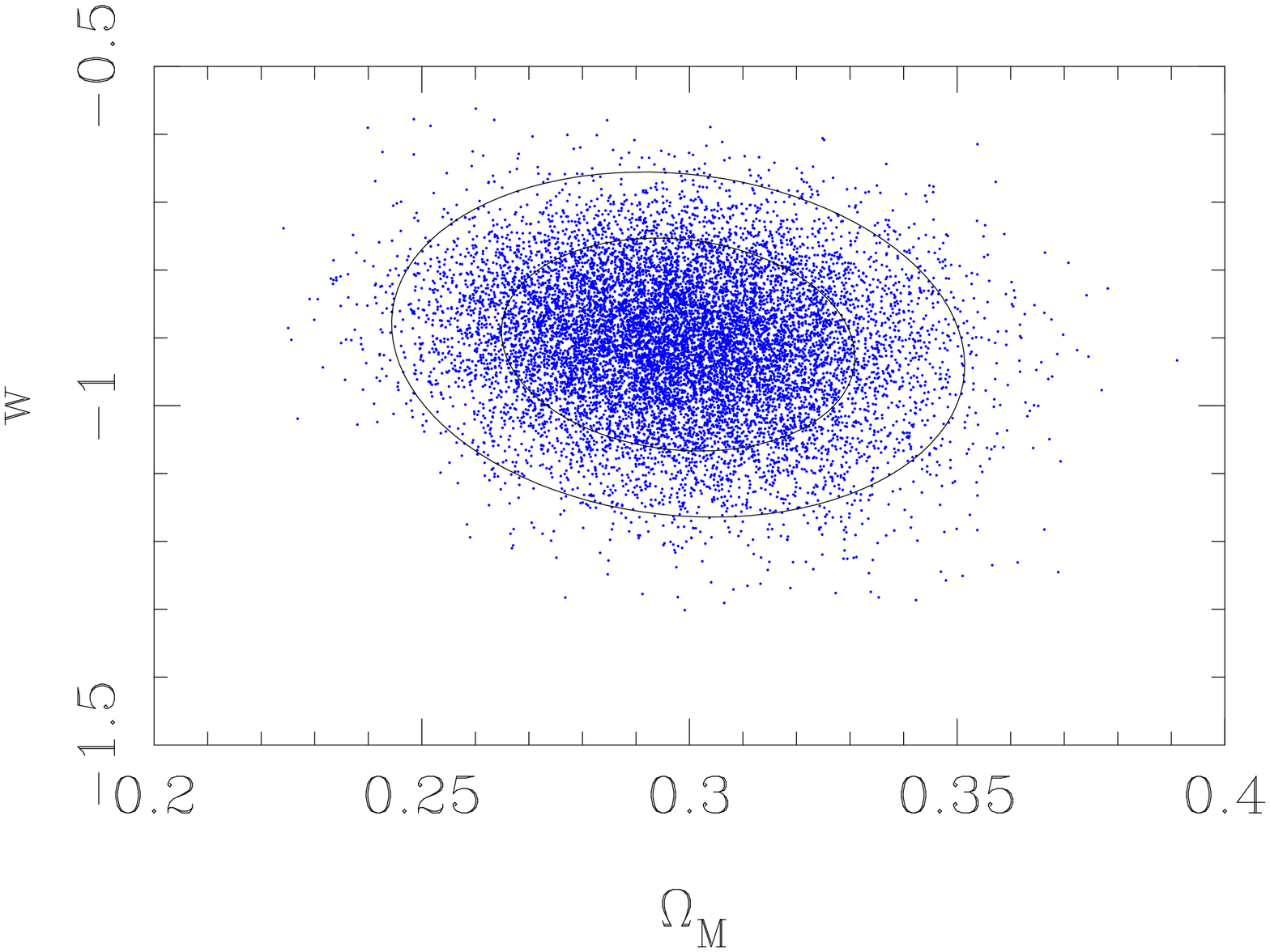,angle=-90,width=0.45\textwidth,bb=130 40 560 620}
   \caption{Examples of the outcome of a SMOCK run. The points show
     the best fit
    values of 8000 simulated data sets similar to the Gold data set. 
    The inner
    and outer contours, obtained from the density of points,
    correspond to 68.3\% and 95.4\% confidence 
    level. Left and right panels show the best fit values in the 
$(w_0,w_a)$-plane and $(\Omega_M,w)$-plane, respectively.
    \label{fig:smocksample}
}
\end{center}
\end{figure}

In step (iii) we use the best fit values of $\vec{\theta}$ from a
large number of realizations of synthetic data-sets to construct
confidence level contours.  The best fit parameters $\vec{\theta}$
obtained for each synthetic data-set can hence be viewed upon as one
possible "measurement" of the cosmological parameters including
systematic uncertainties. The ensemble of best-fit values stemming
from the iteration procedure maps out the confidence region of
possible "measurements" given the data with systematic and statistical
uncertainties. Figure~\ref{fig:smocksample} shows examples of the best
fit values of $w$ and $\Omega_M$ as well as $w_0$ and $w_a$ for 8000
perturbed ``Gold'' data-sets. In each case, the fit also includes a
BAO prior from \cite{eis05}. To obtain the confidence level contours in
Figure~\ref{fig:smocksample} and subsequent figures we use 
the density of best fit values of the cosmological parameters
$\vec{\theta}$.  The contours corresponding to e.g.~the 68.3\%
confidence level in the plane spanned by $\Omega_M$ and $w$ 
(right panel in Figure~\ref{fig:smocksample})
and $w_0$ and $w_a$ (left panel) are found by
identifying the isocontours which enclose 68.3\% of the
points.\footnote{For figures displayed here we have used the elliptic
contour that closest match the true isocontour. The two are generally
similar and elliptic contours make comparisons clearer.} We
have checked that the confidence level contours obtained in this way
agrees with contours from standard $\chi^2$ minimization, when the
errors are Gaussian and uncorrelated. In the standard interpretation
of confidence level contours, the 68.3\% confidence level contour, for
example, is the contour which fulfills $\chi^2=\chi^2_{\rm min}+2.30$.
\footnote{Subtle differences between the two interpretations of confidence level could arrise. So far no deviations have been detected, but a more dedicated study might be of interest and is a possible topic for future SMOCK-related work.}


This iterative method was developed mainly in order to take
correlated and non-Gaussian errors into account (e.g. from 
extinction and gravitational lensing). When using standard
maximum likelihood techniques, all errors are required to be Gaussian
and the correlation matrix has to be calculated and inverted for each
systematic error. When dealing with large  
data-sets this inversion can be numerically cumbersum and extremelly
time-consuming. Using the SMOCK method this is avoided -
each iteration is treated as a single occurrence and only the best fit
is calculated.\footnote{The covariance matrix for any error could in
  principle be calculated through the Monte Carlo simulated
  data-sets.} It should be noted that in this fit [step (ii)] 
we use the standard
procedure of assuming all SNe~Ia to be uncorrelated with Gaussian errors,
but since a) only the best fit is calculated and b) a large number of
individual fits are used the result will be more accurate and the
process less demanding for large 
data-sets.

The code developed to run the loop consisting of steps (i)-(iii), and in
particular to generate simulated data-sets taking systematic effects
into account in a flexible way, is collected in the SMOCK package.

\subsection{An example: the Gold data-set}
As a test case, we have chosen to study the ``Gold'' sample presented
in Reference~\cite{riess06}\footnote{From which we have used SN~Ia redshifts,
  colors, magnitudes and corresponding errors. The colors were derived
  from the host galaxy extinction, $A_V$, through the assumption of $E(B-V) =
  A_V/R_V$ where $R_V = 3.1$.}.  The data-set was further divided
into three primary redshift bins: i) 36 low-z SNe~Ia ($z<0.1$), ii) 90
intermediate redshift SNe~Ia ($0.1<z<0.7$), and iii) 56 high-z SNe~Ia
($z>0.7$). This division in redshift space was done in order to study
the impact from systematic effects applied at different redshifts. 
It should be emphasized that this division of supernovae is
somewhat arbitrary and mostly done for demonstration purposes.
However, the redshift bins broadly correspond to i) the near-by SN~Ia
sample used to anchor the Hubble diagram \cite{Hamuy96,Riess99,jha06},
ii) the ground based data from the High-Z Team, the Supernova Cosmology
Project and the SNLS, and iii) the HST data from the highest-z SNe~Ia.

\section{Propagation of systematic effects}
\label{sec:sysquant}
\begin{figure}[hbtp]
\begin{center}
    \epsfig{figure=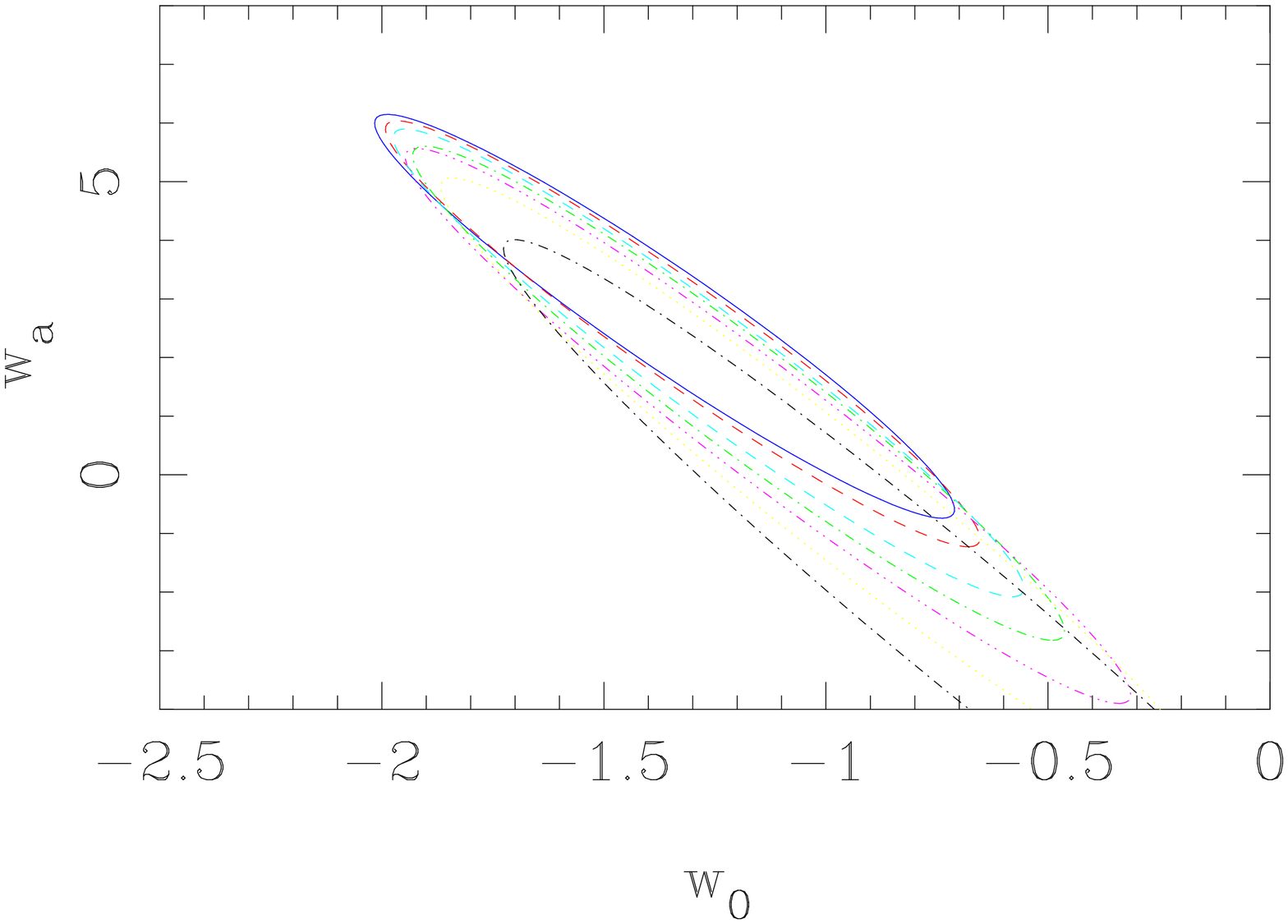, angle=-90,width=0.45\textwidth,bb=130 40 560 640}
    \epsfig{figure=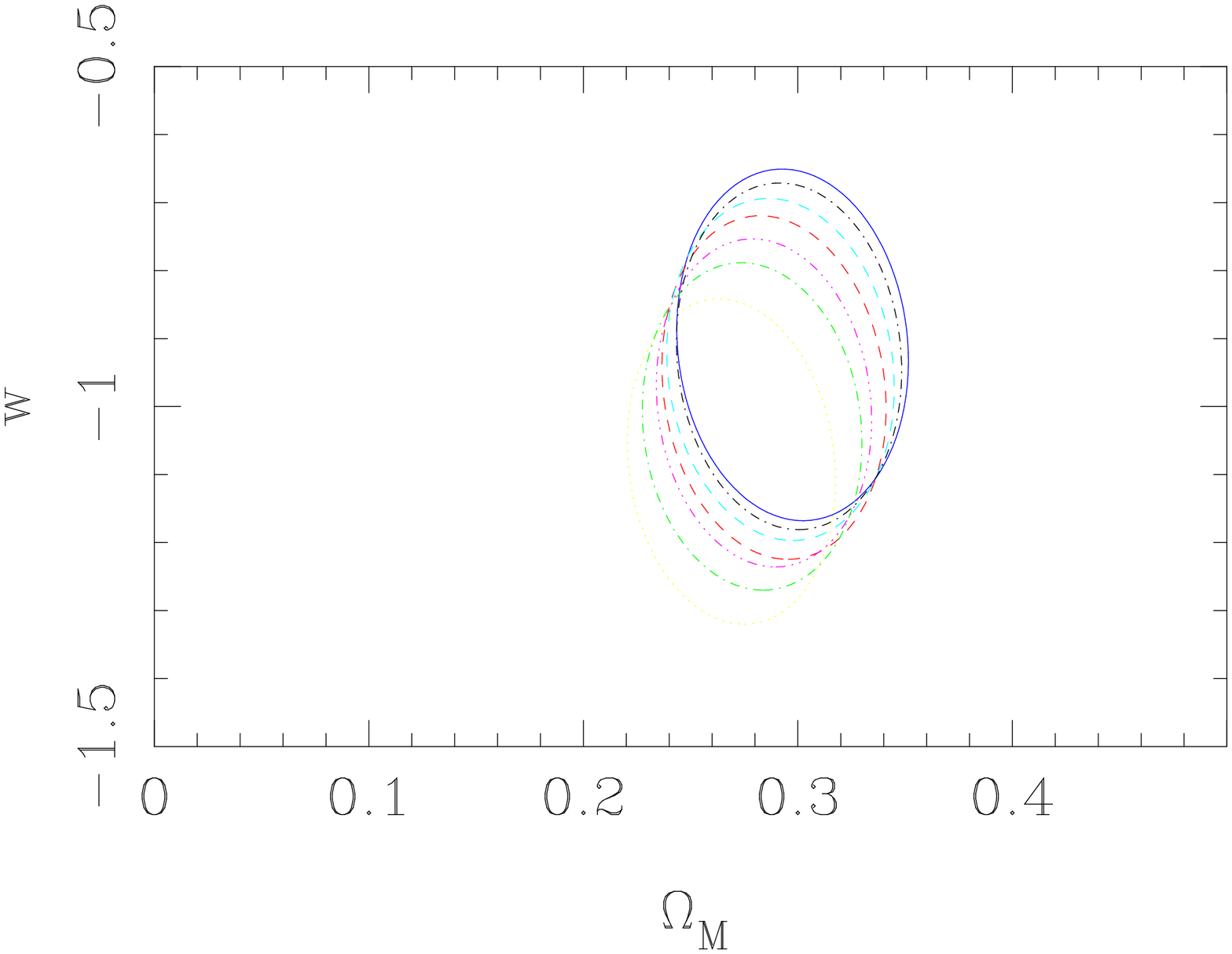, angle=-90,width=0.45\textwidth,bb=130 40 560 620}
    \caption{Effect on cosmological results due to high-z SN~Ia with 
      biased magnitudes. 
      Left and right panels show 68.3\% confidence level contours 
    in the 
     $(w_0,w_a)$-plane and $(\Omega_M,w)$-plane, respectively.
    Solid (blue) contours correspond to
    the results of the clean run, which only take into account 
    statistical uncertainties. The other contours show the effect of
    adding bias of 0.01, 0.03, 0.05, 0.07, 0.1, and 0.15 mag. 
     Flat universe and BAO prior assumed.
    }
    \label{samplefig:highbias_cont}
\end{center}
\end{figure}

Through repeated SMOCK runs using a range of 
errors of different magnitudes we are able to make 
comparisons between 
the impact on fitted cosmological parameters from different 
systematic effects and standard statistical uncertainties. 
Figure~\ref{samplefig:highbias_cont} shows an example of the parameter
fits which would result if the peak brightness magnitudes of the 
high-z sample were positively biased.
The solid contour shows the 68.3\% confidence level contour for the
unperturbed (clean) sample, which only suffers from statistical
uncertainties. The other contours in the figure correspond to data
sets affected by high redshift apparent magnitude bias
in the range $0.01$ to $0.15$ mag.
An archive consisting of figures of \emph{all} systematic effects
individually applied to each redshift bin in the Gold set can be found
at the SMOCK webpage www.physto.se/\~{}nordin/smock.

To further quantify and compare different systematics we have
monitored the functional dependence of the fitted parameters 
$\vec{\theta}$ and some other indicators 
on the assumed systematics. In what follows we will
refer to the shift of the best fitted values or its changing
uncertainty due to increased systematic effects as
the ``evolution'' of a parameter or indicator.
In particular, we will exemplify the 
evolution of five parameters and indicators due to increasing
levels of systematic uncertainties:

\begin{enumerate}
\item[(i)] The evolution of (best fit) $\Omega_M$ and $w$.
\item[(ii)] The evolution of (best fit) $w_0$ and $w_a$.
\item[(iii)] The evolution of the size (area) of the confidence level
  contours\footnote{This area is always used in comparison to the
   area corresponding to the clean sample.}.

\item[(iv)] The evolution of the \emph{total~area}. This is defined as
  the complete confidence region that would have to be used \emph{if}
  the systematic error was suspected.\footnote{This parameter will
    depend on the changes of the first three, as well as the relative
    ``direction'' of error propagation in the fitted parameter plane.
    Looking at Figure~\ref{samplefig:highbias_cont}, an uncertainty of up
    to 0.15 mag would mean that the \emph{total} area covered by
    \emph{any} ellipse plotted should be included in the total area.
  This area is quoted compared to the
   area corresponding to the clean sample. }
\item[(v)] The increased dispersion around the best fit model in the 
Hubble diagram. Possible systematic effects that increase the scatter above some limit would be considered unrealistic.
\end{enumerate}

We have parametrized the evolution  of the five parameters and
indicators listed
above as a function of the magnitude of the systematic effects. The
goodness of each parameterization is estimated through the residual
scatter around the best fit of the evolution.  Notice that these 
parameterizations depend on the data set used and the redshift 
binning chosen. In most
cases a parameterization linear with the size of the systematic effect
yields a good fit. Where quadratic or exponential parameterizations have 
been used, this is noted.

Uncertainties added to the SN~Ia data affect the dispersion in the
Hubble diagram. This provides a way of estimating whether systematic
effects are realistic or not - a systematic effect introducing a
dispersion larger than the observed one is unlikely. We have therefore
also monitored the added {\em dispersion} on the Hubble diagram as
systematic effects were introduced (i.e.~increased residual scatter around the best fit cosmology).  We find
that constant bias often yield very small changes in the original
dispersion, while added scatter give rise to 
increased dispersion, as expected.
In general, the bias from systematic effects begin to
dominate over statistical errors well before causing a noticeable increase
in the residuals in the Hubble diagram. A typical example will be 
discussed below (see~\S\ref{sec:rv}).

As an example of how the SMOCK technique could be used, we calculate
 the limits for when each individual systematic effect will dominate
 over the statistical errors for the Gold data-set. This was defined to
 occur either when the area of the $95.4\%$ confidence level contour
 in the $(\Omega_M,w)$-plane or the $(w_0,w_a)$-plane was doubled
 (compared to the clean fit) or when the best fit of $\Omega_M$ or $w$  
 or, in the other case, $w_0$ or $w_a$
 was outside the $95.4\%$ confidence level region of the clean fit.
 This might be called the point of "systematic-statistic equality",
 and is used by us to compare the relative importance of different
 systematic effects.  These points are summarized in
 Table~\ref{tab:smallbig} (for the $w_0-w_a$ parametrization) 
and Table~\ref{tab:omegasmall} (for the $\Omega_M-w$ parametrization).

\begin{figure}[btp]
\begin{center}
    \epsfig{figure=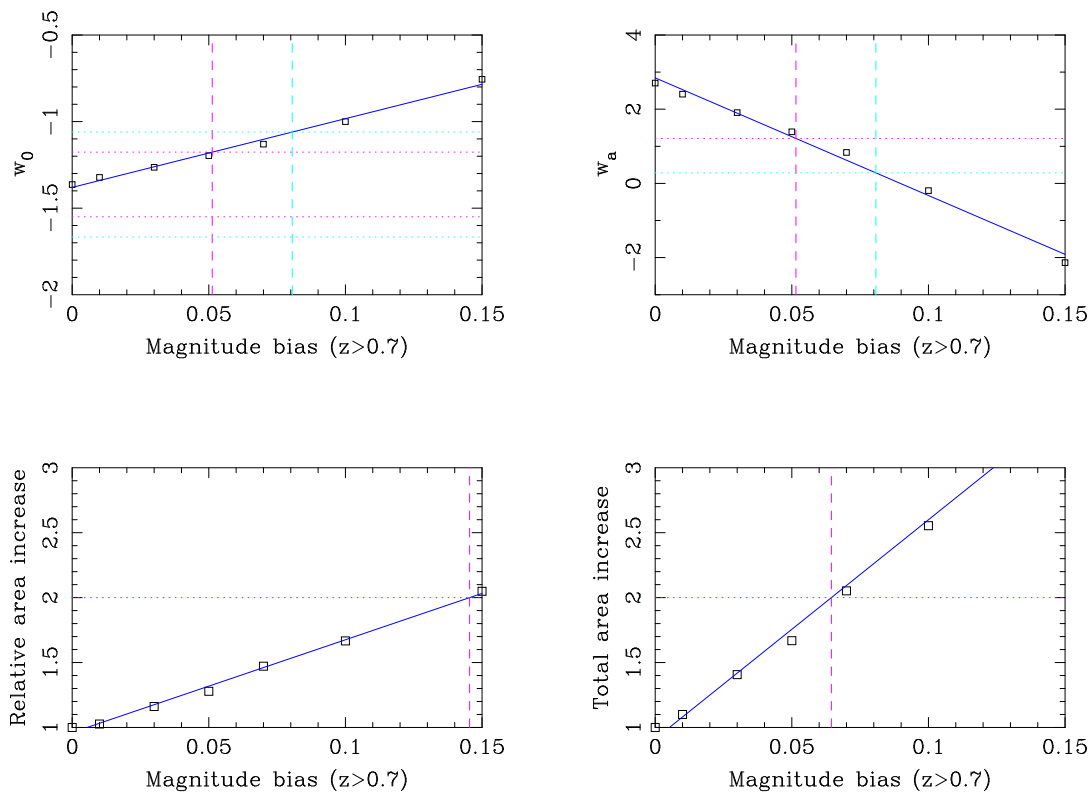, angle=-90,width=0.8\textwidth,bb=51 45 570 657}
    \caption{Magnitude bias for high-z SN~Ia (systematic offset for all SNe with $z>0.7$). Horizontal lines in the 
top left and top right plot show the limits when the statistical 
68.3\% and 95.4\% confidence levels are reached by systematics. 
In the bottom left and bottom right plot the horizontal lines show
when the area relative to the clean run is doubled. Vertical lines
show at which error (if so) these limits are reached. This signifies 
either a bias creating an effect larger than the
statistical uncertainty or a systematic uncertainty doubling the size 
of the regions (or a combination thereof).}
    \label{samplefig:highbias_errs}
\end{center}
\end{figure}

\begin{figure}[btp]
\begin{center}
    \epsfig{figure=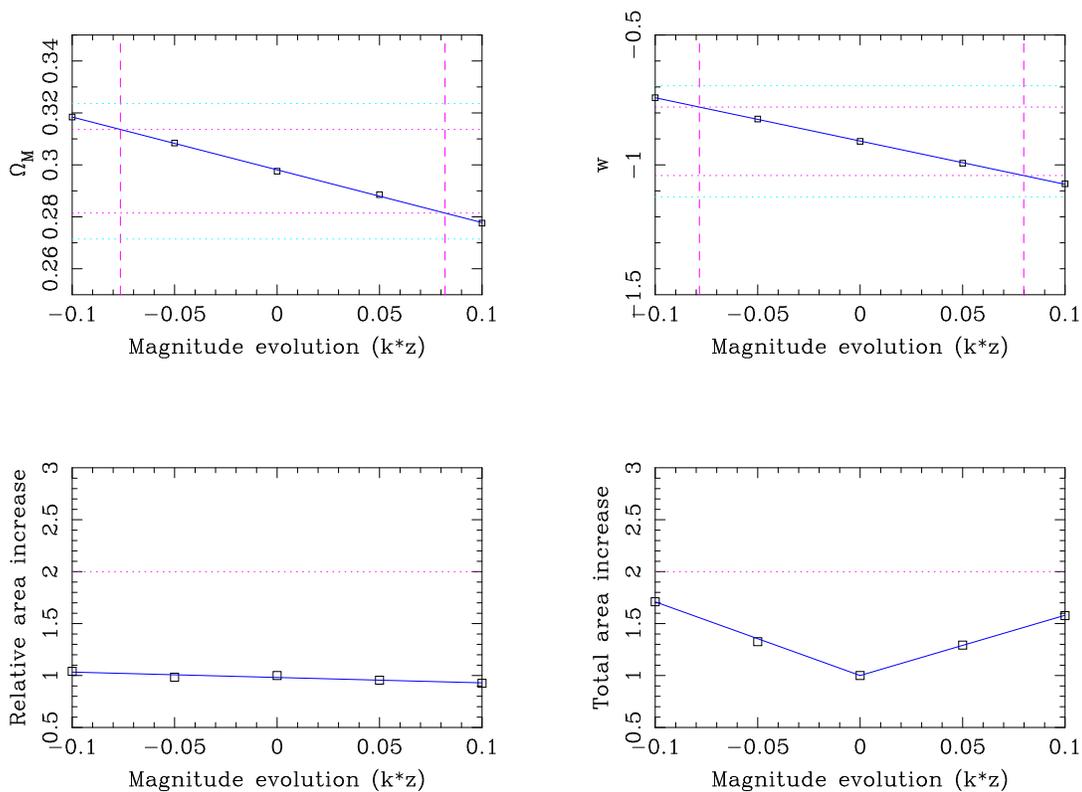, angle=-90,width=0.8\textwidth,bb=51 45 570 657}
    \caption{Bias in the $(\Omega_M,)w$-plane caused by evolution of peak brightness as a function of redshift. Horizontal lines in the 
two top plots show limits when statistical 68.3\% and 95.4\%
confidence levels are reached by systematics. In area plots horizontal lines show
when the area relative to the clean run is doubled. Vertical lines
show at which error (if so) these limits are reached. This signifies
either an introduced bias creating an effect larger than the
statistical uncertainty or an introduced uncertainty doubling the size
of the regions (or a combination thereof).
}
    \label{samplefig:omega_errs}
\end{center}
\end{figure}

Figure~\ref{samplefig:highbias_errs} shows the evolution of the
parameters and indicators described in Section~\ref{sec:sysquant} 
for the Gold sample
with added constant magnitude bias. Figure~\ref{samplefig:omega_errs}
shows the evolution in the $(\Omega_M,w)$ plane when the absolute 
magnitude of SN~Ia evolves with redshift. Graphs showing the evolution of
systematic errors for a larger range of simulated errors can be
found at www.physto.se/\~{}nordin/smock. Some of the main results for the
different systematics are presented below.


\begin{figure}[btp]
\begin{center}
    \epsfig{figure=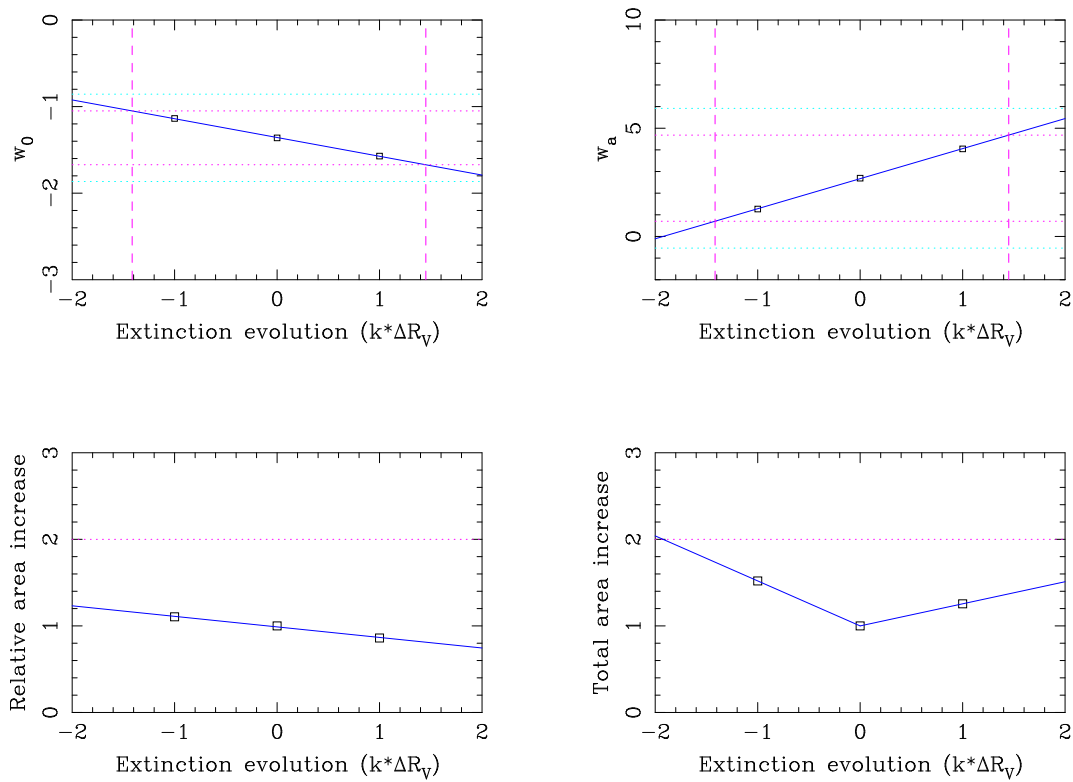, angle=-90,width=0.8\textwidth,bb=51 45 570 657}
    \caption{Effects from dust extinction evolving with redshift parameterized as a linear modification ($\delta R_v = k*z$, $k$ systematic error) of local (Milky Way) values. Effects on $w_0-w_a$ plane demonstrated. Horizontal lines in the 
two top plots show limits when statistical 68.3\% and 95.4\%
confidence levels are reached by systematics. In area plots horizontal lines show
when the area relative to the clean run is doubled. Vertical lines
show at which error (if so) these limits are reached. This signifies either an introduced bias creating an effect larger than the statistical uncertainty or an introduced uncertainty doubling the size of the regions (or a combination thereof).}
    \label{samplefig:grad_errs}
\end{center}
\end{figure}

\Table{\label{tab:smallbig}
Points of systematic-statistic equality for different
 systematics. Parameters and indicators in the table are described in
 Section~\ref{sec:sysquant}. All fits are linear unless otherwise
 noted.}
\br
Error & z-bin & $w_0$ & $w_a$ & Area & Tot. Area \cr
\mr
$E(B-V)$ bias&high&\00.011&\00.011&\m---&\0\m0.015 \cr
$E(B-V)$ bias&int&\00.016&\00.016&\m---&\0\m0.031 \cr
$E(B-V)$ bias&low&\00.015&\00.014&\m---&\0\m0.022 \cr
$E(B-V)$ spread&high&---&---&\0\m0.086$^{\rm a}$&\0\m0.084$^{\rm a}$ \cr
$E(B-V)$ spread&int&---&---&\0\m0.031 &\0\m0.031 \cr
$E(B-V)$ spread&low&---&---&\0\m0.100&\0\m0.098 \cr
Evolution &all&\00.09&\00.09&\0\m0.5&\0\m0.1 \cr
Mag bias&high&\00.051&\00.051&\0\m0.145 &\0\m0.064 \cr
Mag bias&int&\00.064&\00.064&\0$-0.742$&\0\m0.128 \cr
Mag bias&low&\00.058&\00.057&\0\m1.353&\0\m0.098 \cr
Mag spread&high&\02.582&\02.607&\0\m0.398$^{\rm a}$&\0\m0.389$^{\rm a}$ \cr
Mag spread&int&\02.798&\02.731&\0\m0.287$^{\rm b}$&\0\m0.279$^{\rm b}$ \cr
Mag spread&low&---&---&\0\m0.412$^{\rm b}$&\0\m0.408$^{\rm b}$ \cr
Lensing&all&---&---&\0\m4.9\%~inc.&\m---\cr
$R_V$ spread&all&18.433&18.586&\0\m1.132$^{\rm b}$&\0\m1.109$^{\rm b}$ \cr
$R_V$ spread&high&42.159&48.616&\m13.399&\m13.168 \cr
$R_V$ spread&int&\05.548&\03.610&$-11.903$&$-58.987$ \cr
$R_V$ spread&low&\02.006&\01.950&\0$-0.882$&\0$-1.454$ \cr
$R_V$ bias&high&\01.7&\01.7&\m---&\m--- \cr
$R_V$ bias&int&11.0&11.0&\m---&\m--- \cr
$R_V$ bias&low&\01.1&\01.1&\m---&\m--- \cr
$R_V$ bias&all&\00.4&\00.4&\m---&\m--- \cr
Redshift bias&high&\00.045&\00.040&\0$-0.171$&\0\m0.108 \cr
Redshift bias&int&\00.007&\00.007&\0\m0.012&\0\m0.006 \cr
Redshift bias&low&\00.001&\00.001&\0$-0.028$&\0\m0.002 \cr
Photometric $z$&all&\00.006$^{\rm b}$&\00.006$^{\rm b}$&\0\m0.003$^{\rm a}$&\0\m0.003$^{\rm a}$ \cr
Redshift spread&high&\00.141&\00.140&\0\m0.597&\0\m0.166 \cr
Redshift spread&int&\00.068$^{\rm b}$&\00.051$^{\rm b}$&\0\m0.042$^{\rm b}$&\0\m0.035$^{\rm b}$ \cr
Redshift spread&low&\00.008$^{\rm b}$&\00.008$^{\rm b}$&\0\m0.004$^{\rm b}$&\0\m0.003$^{\rm b}$ \cr
$R_V$ Evolution&all&\01.452&\01.450&\m--- &\m--- \cr
Malmquist bias&high&11.4\%&11.4\%&\m---&\m--- \cr
Malmquist bias&int&26.9\%&28.0\%&\m---&\m--- \cr
Malmquist bias&low&71.2\%&75.7\%&\m---&\m50\% \cr
\br
\end{tabular}
\item $^{\rm a}$ An exponential fit to the error level had to be used to reach good precision
\item $^{\rm b}$ A quadratic fit to the error level had to be used to reach good precision.
\end{indented}
\end{table}

\Table{\label{tab:omegasmall}Points of systematic-statistic equality for different
 systematics. Parameters and indicators in the table are described in
 Section~\ref{sec:sysquant}. All fits are linear unless otherwise
 noted.}
\br
Error & z-bin & $\Omega_M$ & $w$ & Area & Tot. Area \cr
\mr
Bias&high&\00.103&\00.100&---&\00.277 \cr
Bias&int&\00.181&\00.179&---&\00.198 \cr
Bias&low&\00.070&\00.069&---&\00.128 \cr
$E(B-V)$ bias&high&\00.024&\00.024&---&\00.067 \cr
$E(B-V)$ bias&int&\00.036&\00.037&---&\00.040 \cr
$E(B-V)$ bias&low&\00.017&\00.017&---&\00.033 \cr
$E(B-V)$ spread&high&---&---&\00.222&\00.214 \cr
$E(B-V)$ spread&int&---&---&\00.380&\00.378 \cr
$E(B-V)$ spread&low&---&---&\00.126&\00.122 \cr
Mag evolution&all&\00.081&\00.080&---&\00.326 \cr
Malmquist bias&high&24.8\%&25.6\%&---&--- \cr
Malmquist bias&int&50.9\%&49.7\%&---&--- \cr
Malmquist bias&low&65.7\%&66.2\%&---&69.4\% \cr
$R_V$ bias&high&\04.0&\04.0&---&--- \cr
$R_V$ bias&int&14.2&12.8&---&--- \cr
$R_V$ bias&low&\01.2&\01.2&---&--- \cr
$R_V$ evolution&all&\03.964&\03.932&---&--- \cr
$R_V$ spread&all&---&---&\03.717&\03.482 \cr
$R_V$ spread&high&---&---&10.650&11.482 \cr
$R_V$ spread&int&---&---&35.443&36.061 \cr
$R_V$ spread&low&---&---&\07.445&\07.172 \cr
Mag spread&high&\00.961&\00.948&\02.869&\01.162 \cr
Mag spread&int&\02.805&\02.511&10.656&\03.012 \cr
Mag spread&low&\00.698&\00.721&\00.803&\00.508 \cr
Redshift bias&high&\00.039&\00.039&---&--- \cr
Redshift bias&int&\00.085&\00.092&---&--- \cr
Redshift bias&low&\00.001&\00.001&---&--- \cr
Photometric $z$&all&\00.007&\00.007&\00.007&\00.005 \cr
Redshift spread&high&\00.644&\00.693&\00.396&\00.296 \cr
Redshift spread&int&\00.032&\00.032&\00.271&\00.042 \cr
Redshift spread&low&\00.007&\00.007&\00.008&\00.005 \cr
\br
\end{tabular}
\end{indented}
\end{table}





\section{Specific sources of systematic errors} \label{sec:syseff}
In the current study, we have concentrated on a limited number of
tests for systematic effects.  We emphasize that the SMOCK code is flexible
enough to easily incorporate further effects.  
In this section we discuss motivations, implementations and general results of
the systematic effects we have studied.

The color correction $A_{XY}$ was modelled according to the Galactic
extinction law in Reference~\cite{fitz98}, with the parameter $R_V =
A_V/E(B-V)$ specifing the wavelength dependence. If an average Milky
Way value of $R_V= 3.1$ is assumed, it is possible to determine $A_V$
(i.e. restframe $V$-band absorption) through the measured supernova
color excess, $E(B-V)$. In studying systematic effects we are not
concerned with the absolute value of the correction $A_{XY}$, but in
its change as any of the parameters $R_V$ and $E(B-V)$ change.

A particular difficulty comes with the use of cross filter K-corrections,
$K_{XY}(z)$, and spectral templates. Applying accurate K-corrections
is notoriously challenging since they rely on
spectral templates from compilations of nearby supernovae.  Obtaining a
spectral sequence as a function of time for each high redshift SN~Ia is
practically impossible with current allocation of instruments for 
follow-up observations. Since the same spectral templates are often
used, any errors in the templates would propagate through the
K-corrections and give rise to correlated errors
between supernovae at similar redshifts.

We have included the effects of possible systematic effects due to
K-corrections and color corrections in two ways: 
\begin{itemize}
\item[(i)] When any of the parameters (e.g.~$z$,
$E(B-V)$ or $R_V$), were perturbed both clean and perturbed
corrections were calculated for each SN~Ia\footnote{These perturbed
  corrections include
  integration of the filter with extincted and redshifted spectra. We have used standard Bessel filters for this
  study and unless otherwise specified a modified Nugent-type spectral
  template~\cite{nob03}. In each individual K-correction calculation,
  the filter most overlapping with the $B$-filter redshifted by a
  factor $(1+z)$ is used. } and the difference was used as an
individual SN~Ia magnitude modification.
\item[(ii)] To study the effects from different spectral
templates we compared the differences in cosmological results which
occur for three different spectral templates used in the calculation of 
the K-corrections. 
\end{itemize}

In order to study variations with redshift, systematic effects
were applied to individual redshift bins. It is 
important to notice that systematic errors, including K-corrections,
usually do not add linearly. For realistic combinations of different 
errors, simulations combining all of these are necessary. A
simple addition of errors due to different systematic effects 
might consequently not correspond to the true combined error.

\subsection{Magnitude bias or dispersion}
Instrumentation errors, calibration errors, peculiar velocities, or
bulk motion could in principle give rise to redshift dependent 
bias or increased dispersion in SN~Ia magnitudes. We have investigated
the effects of magnitude bias and dispersion for different redshifts.
Figure~\ref{samplefig:highbias_cont} shows an example of the effect of
magnitude bias for the high redshift bin.

\paragraph{Results}
An additional Gaussian dispersion (e.g.~due to
measurement errors and intrinsic dispersion) give rise to increased 
confidence level contours and do not introduce any bias, as expected. 
Constant magnitude bias, on the other hand,
induces a roughly linear bias in the cosmological parameter fits. 
Qualitative confirmation of this can be found from the points of
systematic-statistic equality as given in Table~\ref{tab:smallbig}. An
increased magnitude dispersion does not cause any systematic shift 
in $w_0$ or $w_a$. However, a dispersion of 0.40, 0.29 or 0.41
magnitudes added to the low, intermediate, or high redshift
bin, respectively, doubles the size of the confidence regions.
In a similar way a constant magnitude bias of roughly $\Delta M=0.05$ 
(slightly depending on redshift range) 
shifts the position of the $95.4\%$ confidence level contour so that
the best fit corresponding to unbiased data lies outside the contour.

The direction of the resulting bias 
in the $(w_0,w_a)$-plane depends on the
redshift of the biased supernovae.  A constant magnitude bias in the
intermediate-z bin gives rise to a bias roughly along the major 
axis of the confidence level contour in the $(w_0,w_a)$-plane.
A magnitude bias affecting only the high-z bin, on the other hand,
induces a shift roughly along the direction perpendicular to the
intermediate-z case, i.e.~along the semi-major axis of the confidence
level contour. The cosmological results are consequently more
sensitive to a constant magnitude bias at high-z\footnote[1]{ It might be of interest to note that adding a high-z bias of $\Delta M=0.1$ mag
brings the best fit of the Gold data-set from a slight ``offset'' to
quite precisely that corresponding to a cosmological constant,
$(w_0,w_a)=(-1,0)$.  } than at intermediate-z.

Similar trends can be seen in the $(\Omega_M,w)$-plane, although the
fits are roughly twice as stable considering systematic magnitude
modifications (a bias of rougly $\Delta M=0.1$ mag would 
reach statistical-systematic equality).

Notice that these runs do not include the effects on color 
determination that magnitude errors could create 
(see~\S\ref{sec:color} on color effects below).

\subsection{Redshift uncertainties}
The uncertainty in redshift propagates mainly in four ways into the 
estimation of cosmological parameters: 
\begin{enumerate}
\item The lightcurve shape parameter used in the width-brightness 
      relation, e.g.~the ``stretch'' factor $s$ (see ~\cite{perl97}) is 
      degenerate with the redshift as the latter is obtained from a
      fit of $s(1+z)$ to the light curve. An error in $(1+z)$ 
      is thus compensated
      by a corresponding error in $s$.
\item The K-corrections applied correspond to the wrong redshift.
\item The estimate of the color excess is biased.
\item The points are offset in the horizontal axis of the Hubble diagram.
\end{enumerate}

Points (i) and (iv) can be studied either separately or together,
results quoted here include  effects (ii-iv) combined: no
specific light curve shape dependent modifications were included in the
current version of the code, but this is a simple extension 
planned for future studies since $\delta z = \delta s$.

\paragraph{Results} 
As long as spectroscopic redshifts are available either
from the host galaxy or the SN~Ia, the redshift uncertainty would
typically only affect the second or third decimal place of the 
measured redshift. This is the case for most if not all the SNe~Ia
in the Gold sample.
We note that a bias in the redshifts of the 
low-z (int-z) bin at the level of $\Delta z$ $=0.001$ ($\Delta z=0.007$) 
would suffice to exceed the statitsical uncertainty (when considering
the $w_0-w_a$ paramerization). Thus, this source
of error is not negligible, although it is unlikely to 
be a major concern. However,
future large-scale surveys may have to rely upon photometric redshifts
for a large fraction of the supernovae.  
Studies of photometric redshifts~\cite{mobasher04,ilbert06,mobasher07} 
have shown that the uncertainties are proportional to $(1+z)$.
We model the size of the photometric redshift errors using the formula
$\delta z=k(1+z)$, where the size of the uncertainty is described by
the parameter $k$. We have studied the increased uncertainty due to
this fact, when applied to all supernovae, for several different values of
$k$. 
Photometric redshift uncertainties
can give rise to large effects. Notable is that Gaussian 
uncertainties can give rise to significant bias. 
For the whole sample, the point of systematic-statistic  
equality is reached when the 
photometric redshift error is $\delta z =0.006(1+z)$ 
(see Table~\ref{tab:smallbig}).
Since $d'_L$ 
changes rather slowly at high redshift, we expect the
cosmological results to be more sensitive to redshift
uncertainties at low and intermediate redshift than at high redshift.
This is in agreement with what we find. If only the low redshift bin
is affected by photometric redshift errors, the systematic errors
starts to dominate at $k=0.008$. Since this value of $k$ is only
slightly larger than what is needed for the whole sample, we conclude
that spectroscopic redshifts are particularly important for low
redshift 
SNe~Ia. 
The same general trend can be seen in the $\Omega_M-w$ parameterization.

\subsection{Systematic effects on color \label{sec:color}}
Any bias or error in measurements of apparent magnitudes will propagate
to the color determination and hence affect corrections for
extinction. Biased magnitudes due to miscalibration between different
filters could hence result in color errors.
Use of multi-band color estimations would help stabilizing color
systematics, 
to some extent limiting these effects.
Both constant offsets and Gaussian errors in the color excess, $E(B-V)$,
were considered as systematic effects.

\paragraph{Results}
Estimation of cosmological parameters is very sensitive to errors in
colors: Already a bias of 0.02 in $E(B-V)$ applied to any of the
redshift bins would create a significant bias in the
$(w_0,w_a)$-plane. For a significant bias in the $(\Omega_M,w)$-plane, a
bias of 0.03 is sufficient. Since $E(B-V)$ is a
differential calculated from two (or more) filters this shows the
extreme care necessary in photometry and filter calibration. The intrinsic
colors of SNe~Ia are poorly known, especially at high redshift. Thus,
we rank this source of error as a critical one.

\subsection{$R_V$ systematic effects}\label{sec:rv}
$R_V$ values, being the reddening due to interstellar
dust in the host galaxy or a combination of effects including SN~Ia
physics and/or circumstellar dust, are notoriously hard to determine. We have
studied how much offsets or random fluctuations from a fixed Milky
Way-like value of $R_V = 3.1$ affect confidence level contours. 
The results indicate the systematic effects which arise from using 
$R_V = 3.1$ in the calculations while
the \emph{true} value is different. 
 If the average value deviates significantly from current Milky
 Way estimates, a systematic bias will occur in any cosmology fit.

In this case, the systematic effect was
introduced for all SNe~Ia (allowing for the fact that virtually no host
galaxy $R_V$'s are known) as well as to the members of each bin separately.

\begin{figure}[hbtp]
    \epsfig{figure=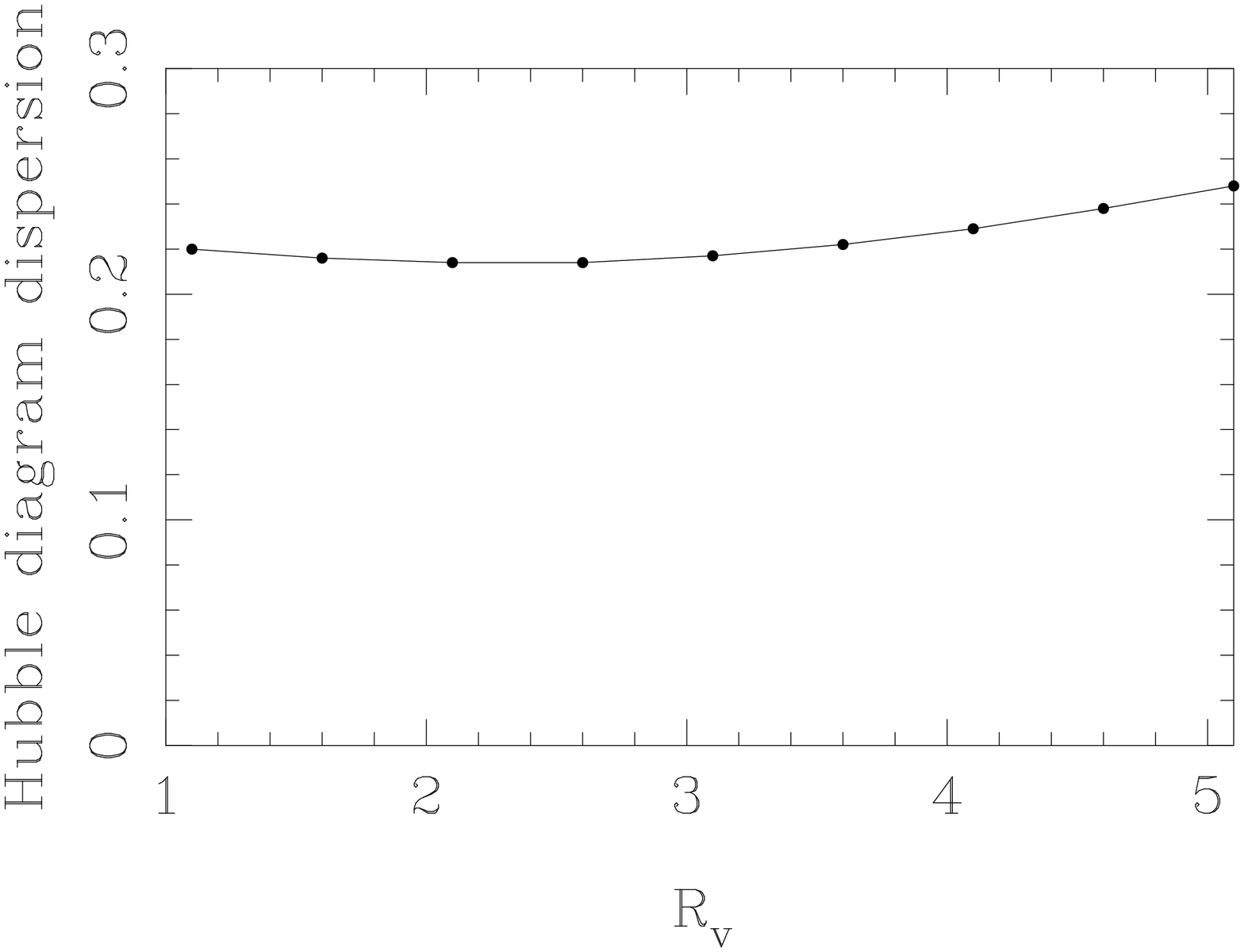,angle=-90,width=0.48\textwidth}
    \epsfig{figure=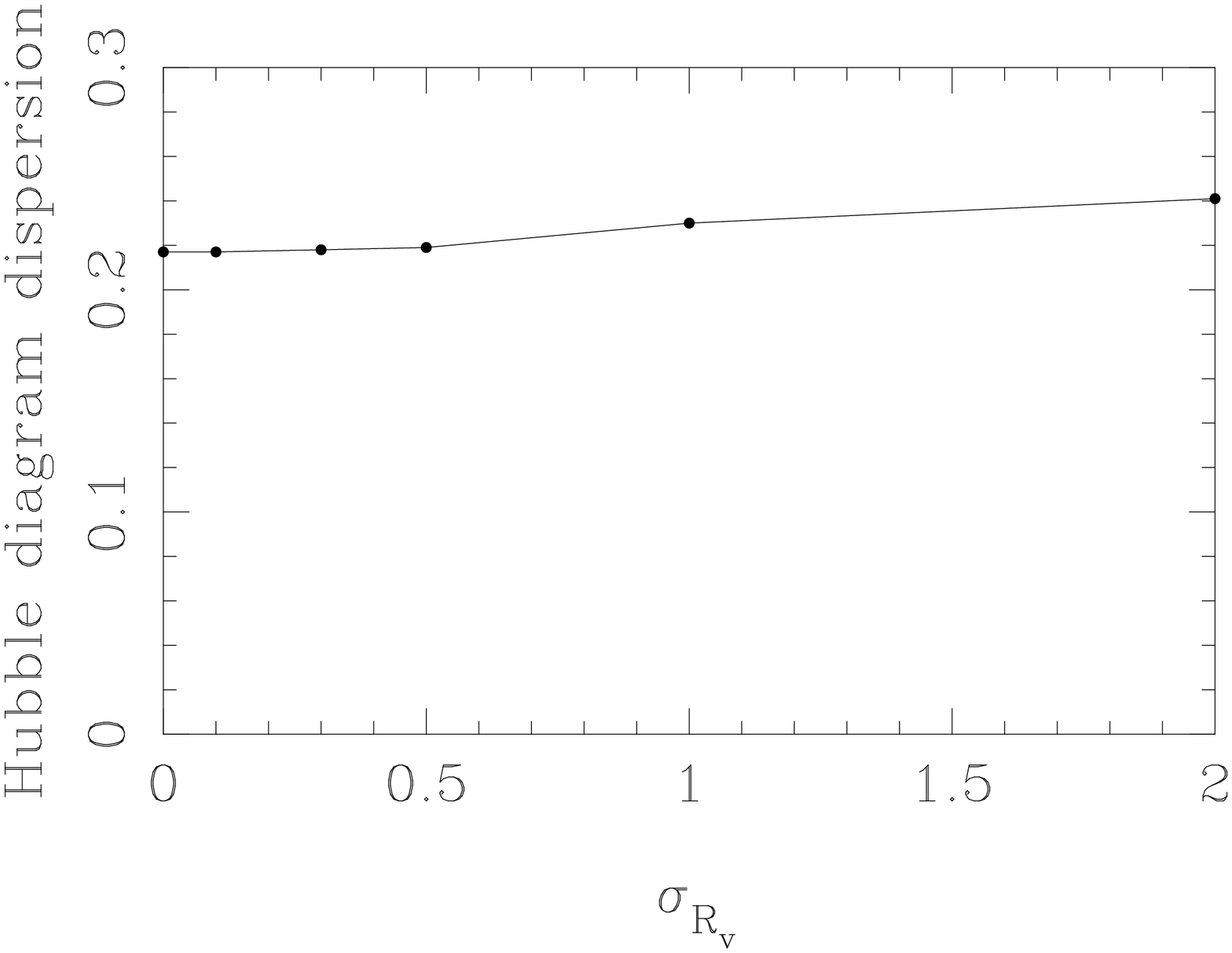,angle=-90,width=0.48\textwidth}
    \caption{\label{fig:hubbledisp} Dispersion (rms) in the Hubble
      diagram due to
    uncertainties in $R_V$ for high redshift SNe~Ia. 
    The right panel shows how the dispersion
    changes for different values of $R_V$. The left panel
    shows the effect of increasing the Gaussian dispersion in $R_V$.
}
\end{figure}

 As an example of how systematic effects influence the scatter in the 
Hubble diagram, consider a bias in $R_V$ for
high-z SN~Ia.  The rms dispersion around the best fit
Hubble diagram for different $R_V$ values is shown in the left
panel in Figure~\ref{fig:hubbledisp}.  The right panel in
Figure~\ref{fig:hubbledisp} shows the dispersion in the Hubble diagram
for different values of the dispersion, $\sigma_{R_V}$, in $R_V$.  
From the figure we
conclude that there is no dramatic increase in the Hubble diagram
dispersion for the range of $R_V$ and $\sigma_{R_V}$
considered.\footnote[2]{A close inspection of Figure~\ref{fig:hubbledisp} reveals a \emph{slightly smaller} dispersion if a small negative $R_v$ bias is added. This shows that care is needed when minimizing hubble diagram residuals since a systematic bias can be introduced.} Other systematics yield similar results.

\paragraph{Results}
As can be seen from Table~\ref{tab:smallbig} a Gaussian with fairly
large width ($\sigma_{R_V}=1.13$) can be added to all SNe~Ia before
reaching the current statistical limit even for $w_0-w_a$ fits. 

But this assumes an $R_V$ distribution which is 
\emph{both} centered on $3.1$ and Gaussian. A non-Gaussian
distribution could 
for example arise from different dust properties in different galaxy
types. Simulations of systematic effects due to non-Gaussian
distributions is another possible future application of SMOCK. 
An offset in the assumed value of total to selective extinction 
coefficient of $\Delta R_V=2$ would be required to significantely
bias the results. We note that this is at the level of discrepancy 
between the Galactic value and what has been derived from SNe~Ia with
the SALT technique \cite{guy05,guy07}. A linear evolution in redshift
$dR_V/dz\approx 1.5$ ($R_V$ evolution in Table~\ref{tab:smallbig}) 
would be required to
exceed the statistical uncertainty in the Gold data-set.

The importance of SN~Ia colors can be seen here: 
For the intermediate redshift bin, the current data-set have average
colors close to zero
making this redshift bin less sensitive to $R_V$-bias. This shows
the advantage of low extinction supernovae and 
in general the promise of how future
subsets of SNe~Ia could be used to limit systematics. 
Obviously, actual knowledge of 
host galaxy dust properties would be preferable.

\subsection{Malmquist bias}
An effect corresponding to Malmquist bias, the systematic failure to detect the faintest elements of a population, was implemented through the option to, in a specific redshift interval, remove a certain fraction of the faintest SNe~Ia and replace them with a (for one SMOCK iteration only) random copy of the \emph{same number} of other SN~Ia in the same redshift interval.\footnote[3]{For a real data-set it might be more relevent to apply a ``reversed'' Malmquist effect, by removing a \emph{negative} fraction of dim SNe~Ia. This will with the current SMOCK-code add new dim objects.} 

\paragraph{Results}
Adding a synthetic Malmquist bias is more natural for a completely
simulated data-set, but runs using the Gold-set does give some
indication of the size of the effect.

Results were quite stable to Malmquist bias added to the data in the low
and intermediate redshift bins. A significant bias did arise 
when Malmquist bias was added to the high-z bin, but in order for
the effect to dominate over the statistical uncertainty a Malmquist ratio 
(the fraction of elements removed) of
about 11\% is required. This would need to be a systematic effect
affecting all SNe~Ia above $z=0.7$. In the $(\Omega_M,w)$-plane the
25\% faintest SNe~Ia 
would have to be missed by a survey to reach the statistical error limit.

\subsection{Gravitational lensing}
Gravitational lensing due to matter along the line of sight to a
supernova, could give rise to amplification or de-amplification of the
observed flux. The effect of gravitational lensing can
be described by the magnification factor $\mu$. Due to flux
conservation the average flux from a large number of standard candles 
is the same as
the flux from a standard candle not affected by lensing. The average
magnification factor is consequently $\langle \mu \rangle =1$. A standard
candle brighter or fainter than average would be described by $\mu >
1$ and $\mu < 1$, respectively. Gravitational lensing increases the
dispersion of standard candles and could give rise to a selection
bias, since amplified SN~Ia are easier to detect than de-amplified ones.
The more distant the standard candle, the more likely it is to
be affected by gravitational lensing since the light emitted from it 
traverses more matter. 
When gravitational lensing is included, all SN~Ia
magnitudes are modified by a term $-2.5\log_{10}\mu$, where the
magnification factors are drawn
from appropriate redshift dependent 
probability distribution functions generated using the SNOC 
package~\cite{snoc}.

\paragraph{Results} 
Added lensing uncertainties does not significantly change the
cosmological results, in accordance with earlier estimates~\cite{jj05}. 
This is an example of how detailed studies of possible
systematics can lead to a situation where a systematic uncertainty can
be said to be under control.

\subsection{SN~Ia spectral templates}
We have compared the results obtained using different spectral templates 
when calculating the K-corrections.
The
data were refitted using both our standard modified Nugent
template~\cite{nugent02,nob03}
and two other publicly available templates~\cite{guy07,hsiao07}. 
This refitting procedure includes new extinction
estimates and new K-correction calculations. The effects of different
templates were studied using two modes: (i) Either the deviation of 
magnitude at epoch 0 (the date of maximum $B$-band brightness) due to 
template differences from our standard template was used or 
(ii) the \emph{maximal} deviation at either epoch $-10$, $0$ or $+10$
days was taken as the systematic error.

\paragraph{Results} 
Since the different spectral templates are often used together with
various light\-curve fitting techniques, the comparisons we report
here are somewhat oversimplified.

Template differences using only epoch 0 values, mode (i), was found
not to be dominating. But in mode (ii) serious systematic shifts do
occur.  The cosmological fits 
performed using the maximal difference between the templates, roughly
differ by 2$\sigma$.
While not being a totally realistic comparison, it
shows the key importance of using accurate SN~Ia templates. These plots are
available through the webpage and the SMOCK Gui (see~\S\ref{sec:gui} 
below).

\subsection{SN~Ia brightness evolution}
It is quite likely that the luminosity of the supernova is related to
the properties of the white dwarf progenitor star, its companion and
details of the propagation of the explosion. On the progenitor side,
key parameters are the mass of the exploding white dwarf, the
relative amounts of carbon and oxygen and metallicity. Thus, evolution
in any of these parameters is a source of concern as it would
lead to a drift in the intrinsic luminosity used to derive distances.
In order to mimic an evolution of this kind, the effects of a redshift
dependent bias were studied. This bias was described by a simple linear
function of redshift, $\Delta M=kz$, where the slope $k$ can be positive or 
negative. 

\paragraph{Results}
Positive or negative evolution of intrinsic SN~Ia brightness with
 redshift produce mainly bias effects. In general, an evolution of 
$|k| \sim 0.1$ is needed to cause significant bias. We show the
 effects 
of evolution in the $(\Omega_M,w)$-plane in
 Figure~\ref{samplefig:omega_errs}, where 
a clear bias in both $\Omega_M$ and $w$ can be seen (as expected, the relative
 area stays the same).

\section{Discussion}\label{sec:disc}
Concerning our present knowledge of dark energy parameters, 
it is plausible that current data
sets of SNe~Ia are plagued by significant systematic uncertainties,
such as cross-calibration errors, spectral template flaws, and uncertain
extinction corrections. Systematic uncertainties may already
dominate the error budget and will certainly do so
in the future. To investigate, understand, and ultimately 
reduce or correct for these sources of error are hence of
uttermost importance for the success of future precision cosmology. 

The purpose of this paper is twofold: to present a method to
investigate the impact on the estimation of cosmological parameters
due to a number of systematic effects and, as an example, investigate
the potential impact of systematic uncertainties in the 
cosmology fits of the Gold data-set of Riess \etal 2007.

The SMOCK method, and all results for the Gold sample presented here, 
are somewhat dependent on the actual redshift binning used 
for the investigation and the details of the
data set. Future improvements would include some form of
parameterization of these effects as well.

In general, any future reported deviation from a
$\Lambda$CDM-cosmology  would need to be supplemented with excellent 
control of all systematics discussed here, since most of them could
account for such a result.

\subsection{Nightmare scenarios}
Various ``nightmare'' scenarios can easily be constructed out of the
systematic effects considered here. Miscalibration between low and
high-z data and biased colors would lead to large systematic effects.

For example, any individual application of either an offset in inrinsic colors $\Delta (B-V)\sim 0.02$,
a linear evolution of the 
lightcurve shape-brightness corrected peak brightness of SNe~Ia exceeding
$\Delta M=0.1 \cdot z$ or
a mean value of $R_V$ of $2$, instead of $3.1$, would be enough to breach the
statistical limit for the Gold data set.

Should any of these effects be combined or added to template uncertainties, systematic effects several times current statistical errrors \emph{could} be generated. 

\subsection{DETF}

Comparing any results presented here with the targets set by the Dark
Energy Task Force (DETF)\cite{DETF} for future (``stage four'') instrument
targets, it is clear that statistical errors will decrease
significantly (as $1/\sqrt{n}$ where $n$ is the number of SNe~Ia). But
systematic effects will not scale in the same way, meaning that the
points of systematic-statistic equality for \emph{bias} studies also
will scale as $1/\sqrt{n}$. Future surveys will thus be
\emph{extremely} sensitive to systematic effects, and \emph{any} bias
will most likely come to dominate over statistical errors.

Thus, to make use of data from fourth generation instruments, control over
instrumentation and calibration issues must be excellent. At the same
time we can hope for a significant increase in the knowledge of
understanding of systematic effects of astrophysical origin
(developments are not easy to predict and are often not included in
instrumentation plans but might be essential for future surveys).

\subsection{SMOCK GUI }\label{sec:gui}
In order to help visualize the effects on the estimation of
cosmological parameters due to systematic uncertainties, we have developed
a small graphical application.
This Graphical User
Interface (GUI)
displays the results obtained by the simulations performed with SMOCK.
The application combines all results obtained for a particular data-set
and allows different combinations of errors to be tried out. This tool, 
together with some of the data obtained for the Gold data-set, 
is available at www.physto.se/\~{}nordin/smock.

It should be noticed that all errors were calculated
\emph{individually}. 
In particular, a combined error in redshift and color-determination
often leads to larger errors (e.g.~through increased K-correction
errors) than those displayed by the GUI.

\section{Conclusions}
\label{sec:concl}
SMOCK, a statistical Monte Carlo simulation approach towards
quantification of systematic effects is presented. This method can
give concrete answers to how sensitive SN~Ia samples are to a wide
variety of different systematic effects. We
can estimate \emph{both} systematic and statistical uncertainties
in a time efficient way.

Current SN~Ia data-sets are likely to be influenced by unknown
systematics, which quite possibly already are larger than the
statistical errors.  In order for future SN~Ia surveys to be
successful, the propagated effects into the cosmolgical fits
 must be understood. With the methods
presented here this will be possible through a process where we first
define acceptable systematic uncertainties to achieve some target
confidence level in e.g. the impact on the derived dark energy properties.
The SMOCK technique can be used to further define the instrumental
requirements necessary to detect and possibly correct for any 
putative source of systematic uncertainty.

Studies of a range of different systematics were performed on the
Gold
data-set, demonstrating the possibility of quantifying sensitivity to 
systematic effects. The importance of excluding the effects of a possible $R_V$ or color bias and obtaining a good spectral template were highlighted. For increased precision future surveys will need to demonstrate that maximal systematic color effects are at the $0.01$ magnitude level and that host galaxy dust properties do not have any systematic shifts over the $\Delta R_V = 1$ level (or that possible such shifts are under control).

\section*{Acknowledgments}
AG would like to acknowledge support by the Swedish Research Council
and the G\"oran Gustafsson Foundation for Research in Natural Sciences
and Medicine.


\vspace{1cm}
\begin{thebibliography}{99}
\bibitem{goo95} Goobar A and Perlmutter S,
 1995 {\it Astrophys. J.} {\bf 450} 14

\bibitem{garnavich98} Garnavich P \etal, 
1998 {\it Astrophys. J.} {\bf 509} 74

\bibitem{riess98}Riess A \etal,  
1998 {\it Astron. J.}  {\bf 116} 1009

\bibitem{schmidt98}Schmidt B~P, \etal 
1998 {\it Astrophys. J.} {\bf 507} 46

\bibitem{perl99} Perlmutter S \etal,
 1999 {\it Astrophys. J.} {\bf 517} 565

\bibitem{knop03} Knop R~A \etal, 
2003 {\it Astrophys. J.} {\bf 598} 102

\bibitem{tonry03} Tonry J \etal 
2003, {\it Astrophys. J.} {\bf 594} 1

\bibitem{barris04} Barris B \etal,
 2004 {\it Astrophys. J.} {\bf 602} 571

\bibitem{riess04} Riess A \etal, 
2004 {\it Astrophys. J.} {\bf 607} 665

\bibitem{astier06} Astier P \etal, 
2006 {\it Astron. Astrophys.} {\bf 447} 31

\bibitem{wood} Wood-Vasey W~M \etal, 
2007, {\it preprint} astro-ph/0701041

\bibitem{riess06} Riess A \etal, 
2007 {\it Astrophys. J.} {\bf 659} 98

\bibitem{neill07} Neill J~D, Hudson M~J and Conley A, 
2007 {\it preprint} arXiv:0704.1654

\bibitem{kcorr} Kim A, Goobar A, and Perlmutter S, 
1996 {\it PASP} {\bf 108} 190

\bibitem{eis05} Eisenstein D~J \etal,
 2005 {\it Astrophys. J.} {\bf 633} 560

\bibitem{powell} Press W~H, Teukolsky S~A, Vetterling W~T, and 
Flannery B~P,
1992 {\it Numerical Recipies in FORTRAN. The Art of Scientific
  Computing}
(Cambridge:University Press, 2nd ed.) p~406

\bibitem{snoc} Goobar A \etal, 
2002, {\it Astron. Astrophys.} {\bf 392} 757


\bibitem{Hamuy96}
  Hamuy M, Phillips M~M, Suntzeff, N~B, \etal, 
1996 {\it Astron. J.} {\bf 112} 2438

\bibitem{Riess99}
  Riess A~G, KirshnerR~P, Schmidt B~P, Jha S, \etal.
  1999 {\it Astron. J.} {\bf 117} 707

\bibitem{jha06}
  Jha S \etal,
2006 {\it Astron. J.} {\bf 131} 527

\bibitem{fitz98} Fitzpatrick E, 
1999 {\it PASP} {\bf 111} 63


\bibitem{perl97} Perlmutter S \etal,
 1997 {\it Astrophys. J.} {\bf 483} 565


\bibitem{mobasher04} Mobasher B \etal, 
2004 {\it Astrophys. J.} {\bf 600} L167

\bibitem{ilbert06} Ilbert O \etal, 
2006 {\it Astron. Astrophys.} {\bf 457} 841


\bibitem{mobasher07} Mobasher B \etal, 
2007 {\it Astrophys. J. Supp.} {\bf 172} 117

\bibitem{jj05} J\"onsson J \etal, 
2006 {\it Astrophys. J.} {\bf 639} 991


\bibitem{nugent02} Nugent P, Kim A, and Perlmutter S, 
2002 {\it PASP} {\bf 114} 803  

\bibitem{nob03} Nobili S \etal, 
2003 {\it Astron. Astrophys.} {\bf 404} 901

\bibitem{guy05} Guy J \etal, 
2005 {\it Astron. Astrophys} {\bf 443} 781

\bibitem{guy07} Guy J \etal, 
2007 {\it preprint} astro-ph/0701828

\bibitem{hsiao07} Hsiao E \etal, 
2007 {\it preprint} astro-ph/0703529


\bibitem{pods06} Podsiadlowski P \etal, 
2006 {\it preprint} astro-ph/0608324


\bibitem{DETF} Albrecht A \etal, 
2006 {\it preprint} astro-ph/0609591

\end{thebibliography}
\end{document}